\documentclass[12pt]{dalthesis}
\pdfoutput=1

\usepackage{amsmath, amsthm, amssymb, epigraph}
\usepackage{enumitem}
\usepackage{xcolor}
\usepackage[hyphens, obeyspaces]{url}
\usepackage{hyperref}
\usepackage[hyphenbreaks]{breakurl}
\usepackage[numbers]{natbib}
\hypersetup{colorlinks=true, linkcolor=blue, urlcolor=blue, breaklinks=true, citecolor=violet}
\usepackage[utf8]{inputenc}
\usepackage{fancyvrb}
\theoremstyle{definition}
\newtheorem{theorem}{Theorem}[section]
\newtheorem{definition}[theorem]{Definition}
\newtheorem{proposition}[theorem]{Proposition}
\newtheorem{lemma}[theorem]{Lemma}

\PassOptionsToPackage{obeyspaces}{url}

\DeclareUnicodeCharacter{211D}{$\mathbb{R}$}
\DeclareUnicodeCharacter{2115}{$\mathbb{N}$}
\DeclareUnicodeCharacter{2261}{$\equiv$}
\DeclareUnicodeCharacter{220E}{$\qed$}
\DeclareUnicodeCharacter{211A}{$\mathbb{Q}$}
\DeclareUnicodeCharacter{1D58}{$^u$}
\DeclareUnicodeCharacter{2262}{$\not\equiv$}
\DeclareUnicodeCharacter{2223}{$\mid$}
\DeclareUnicodeCharacter{2264}{$\le$}
\DeclareUnicodeCharacter{2243}{$\simeq$}

\begin{document}

\title{Constructive Analysis in the Agda Proof Assistant}
\author{Zachary Murray}

\degree{BSc Honours in Mathematics and Computer Science}
\degreeinitial{BSC-MTCS-H}
\faculty{Science}
\dept{Department of Mathematics & Statistics}
\defencemonth{April}\defenceyear{2022}

\nolistoffigures
\nolistoftables

\frontmatter

\begin{abstract}
Proof assistant software has recently been used to verify proofs of major theorems, yet even the libraries of some of the most prominent proof assistants lack much of undergraduate mathematics. In particular, the Agda proof assistant has no formalization of the real numbers and their arithmetic. In this thesis, I present my implementation of Errett Bishop’s constructive real numbers in Agda, including their arithmetic, ordering, and fundamental results, such as uncountability and Cauchy completeness. We will also survey the basic concepts of constructive analysis and the Agda proof assistant.

\end{abstract}

\begin{acknowledgements}
First, thank you to my supervisor, Dr. Peter Selinger, for your guidance and comments, especially on efficiency and design, and for teaching me Agda in the first place. Thanks to Dr. Frank Fu for his teaching of MATH 2112 in the summer of 2018, in particular for briefly banning the use of the law of the excluded middle, which is what led me to study constructivism and type theory. I am also grateful to the David and Faye Sobey Foundation for helping make this project possible with their generous research scholarship during the summer of 2021.
 
\end{acknowledgements}

\mainmatter

\chapter{Introduction}
\section{Informal Mathematics}
Modern mathematicians write their proofs in an informal style in natural language. A paper's readers are assumed to have much expertise in the topic being studied. This allows the paper's authors to skip seemingly trivial details in their proofs, focusing only on the bigger details, and the bigger picture. Informal mathematics, in this sense, is extremely efficient. It is relatively easy to write a proof informally. Likewise, it is relatively easy for the reader to pick up the key ideas and concepts.

However, it can also be easy to make errors in informal mathematics. While it is common for papers written in the informal style to have minor errors, these errors are usually easily fixed. On the other hand, severe errors, which either cannot be fixed or would be very difficult to fix, are relatively uncommon.

But severe errors do appear in informal mathematics, even within published papers. Fermat's last theorem is infamous for the number of erroneous proofs published. Even Andrew Wiles's original proof contained a severe flaw, which took years to fix.

\section{Mathematics Formalized}

Proofs written in a formal setting, like Fitch-style natural deduction, have been a staple of courses in discrete mathematics for years. Such sterilized formalism familiarizes students with the exact logical rules mathematicians are allowed to use in informal proofs. These logical rules are, of course, usually unmentioned in informal proofs, but they are always hidden in the background.

Anyone who has written a fully formal proof, even of simple logical laws, knows it is incredibly time consuming. There is so much detail that the intuition behind the proof is lost to the reader. The benefit is that errors are easily identified. Indeed, it seems such errors could be checked by a computer.

\section{Mathematics Computerized}

If all informal proofs were written out formally so that a computer could read the formal proof and verify it, then severe errors would almost never be published. But writing out such a formal proof is too time consuming, and we first need to program the computer to be able to verify formal proofs.

Proof assistant software aims to solve this problem. Proof assistants allow users to write their proofs somewhere in between informal and formal. Informal enough that, hopefully, anyone reading the proof can pick out the key ideas, and also so that the proof does not take as long to write as a fully formal proof. But formal enough that the proof assistant can figure out the rest of the details, thus constructing and verifying a fully formal proof.

Proof assistants have already been used to eliminate the possibility of error in famous proofs. A proof of the four colour theorem was verified in the Coq proof assistant by Georges Gonthier, finally ending a long history of false proofs and controversy \cite{gonthier}.

\section{Constructive Analysis in the Agda Proof Assistant}

This thesis describes an implementation of mathematics in a proof assistant. We implement a basic theory of real numbers and sequences and series. In particular, we implement these aspects from the American analyst Errett Bishop's constructive analysis. The proof assistant we use is Agda.

In implementing this constructive analysis, we hope to achieve two goals.

\begin{enumerate}
	\item Implement a theory of real numbers in Agda.
	\item Study the problems a user implementing everyday mathematics may encounter in a proof assistant.
\end{enumerate}

For the first goal, we note that, at the time of writing, the Agda standard library is lacking a theory of real numbers. We hope that this thesis will help solve this problem.

For the second goal, it must be admitted that most mathematicians do not use proof assistants. Those that do tend to focus on mathematics other than analysis. It will be useful, then, to see how feasible analysis is in Agda.

The result of this project is an Agda library consisting of over 6000 lines of code. In this thesis, we provide some background in both constructive analysis and Agda so that we can discuss the stated problems.

This thesis is aimed at three audiences. The first audience is the proof assistant community. This audience will probably be more interested in the first problem, and thus in the library itself. For them, the most useful parts of this thesis are probably the background material on constructive analysis in Chapter \ref{constructiveanalysis}, the discussion of the library in Chapter \ref{constructiveanalysisinagda}, and the library at \cite{libraryGit}.

The second audience is mathematicians curious about proof assistants, but who are unsure about how useful they can be and what problems the average user will encounter. This audience will probably care more about the first problem than the second problem. The most useful material for them should be the introduction to Agda in Chapter \ref{Agda}.

The third audience is those interested in learning about constructive mathematics. On the one hand, constructive mathematics, historically done on paper, is increasingly concerned with proof assistants. On the other hand, the design of proof assistants tends to favour constructive mathematics, and this itself is worth studying. Moreover, Bishop's constructive analysis has been an area of keen interest in constructive mathematics since its appearance in 1967. As we will discuss the history and problems of each of these points, every chapter except, perhaps, Chapter \ref{constructiveanalysisinagda}, should be of interest to the third audience.

\chapter{Constructive Mathematics}
\section{History}
Constructive mathematics finds its roots in philosophical debates concerning the foundations of mathematics during the late 19th and early 20th centuries. Questioning the role and validity of infinities in proofs and driven by concerns about set-theoretic antinomies like Russell's paradox, philosophers and mathematicians debated the existence and character of a foundation of mathematics. Thus three competing schools of thought emerged, namely the formalism of German mathematician David Hilbert, the logicism of the logicians Gottlob Frege and Bertrand Russell, and finally, the intuitionism of Dutch topologist L.E.J. Brouwer.

Of the three, it is Brouwer's intuitionism that is constructivist. Intuitionists conceive of mathematics as part of human consciousness. Mathematical objects cannot be said to exist outside of the human mind, and a proposition is proved when it is witnessed by a sequence of mental constructions. From intuitionism, we abstract the core tenet of constructive mathematics: To exist is to be constructed.

As a result, constructivists must not assume the principle of double negation (DNE) (i.e., $\lnot\lnot P\implies P$) since we could then prove $\exists xP(x)$ from $\lnot\lnot\exists xP(x)$. Such proofs do not specify how to construct the $x$ it claims to exist. If we commit to the common view that mathematical objects exist, then DNE forces us to conclude that $x$ must have some sort of external and mind-independent existence. Because the law of the excluded middle (LEM) (i.e., $P\lor\lnot P$) is logically equivalent to DNE, constructivists also reject LEM.

Intuitionism, however, was seen as an untenable foundational philosophy of mathematics. The rejection of LEM places limitations on analysis, and fundamental results like the Intermediate Value Theorem are not provable constructively. Further, the intuitionists subscribed to Kant's view that time is a part of the mind which makes our conscious experience possible, and not part of the external world, and they also rely on the concept of free will. The intuitionistic continuum is dependent on these philosophical imports, and philosophical arguments become vital to analysis. This results in strange conclusions, such as every function of a real variable being continuous. Intuitionism is thus at odds with standard mathematical practice thanks to its philosophical baggage and odd results.

Because intuitionism was the dominant constructivist school, its mystic character became associated with constructivism more generally. Due to the oddities of intuitionistic analysis, it was believed that analysis could not be carried out constructively. Hilbert went as far as to claim that ``Taking the principle of excluded middle from the mathematician would be the same, say, as proscribing the telescope to the astronomer or to the boxer the use of his fists'' \cite[p.~476]{hilbertLEM}.

\section{Intuitionistic Logic}
Modern constructivists use \textit{intuitionistic logic} in place of classical logic. Intuitionistic logic is essentially classical logic without the law of the excluded middle or any equivalent axiom. Perhaps the most popular form of intuitionistic logic is the Brouwer-Heyting-Kolmogorov (BHK) interpretation. A proposition $P$ is identified with the set of its proofs. The interpretation of the logical connectives is summarized in Table \ref{tab:BHKTable}. The left column represents the proposition $P$ we wish to prove, and the right column defines what a proof of $P$ is.
\begin{table}[tbh]
\centering
\begin{tabular}{|p{4cm}|p{10cm}|}
	\hline
	Proposition & Proof Requirement
	\\
	\hline
	$P \implies Q$ & A function mapping proofs of $P$ to proofs of $Q$ \\
	\hline
	$P \land Q$ & A pair $(p,q)$ of a proof $p$ of $P$ and a proof  $q$ of $Q$ \\
	\hline
	$P \lor Q$ & A pair $(0,p)$, where $p$ is a proof of $P$ or a pair $(1,q)$ where $q$ is a proof of $Q$ \\
	\hline
	$\exists xP(x)$ & A pair $(a,p)$ of some $a$ and a proof of $P(a)$ \\
	\hline
	$\forall xP(x)$ & A function that maps objects $x$ to a proof of $P(x)$ \\
	\hline
	$\lnot P$ & A proof of $P \implies \bot$ \\
	\hline
\end{tabular}
\caption{Intuitionistic characterization of logical connectives.}
\label{tab:BHKTable}
\end{table}

\chapter{Constructive Analysis}\label{constructiveanalysis}
\section{History}
\setlength{\epigraphwidth}{\textwidth}
\epigraph{We are not interested in properties of the positive integers that have no descriptive meaning for finite man... If God has mathematics of his own that needs to be done, let him do it himself.}{Errett Bishop, \textit{Constructive Analysis} \cite[p.~5]{bishopBridges}}

Errett Bishop (1928\textendash 1983) was an American analyst and a leading figure in the development of modern constructive mathematics. Like the intuitionists, Bishop believed that mathematics was a creation of the mind, and was disturbed by the free use of the law of the excluded middle and other nonconstructive principles, referring to classical mathematics as ``idealistic'' \cite[pp.~4-12]{bishopBridges}. However, Bishop was disappointed by the intuitionists' focus on philosophical speculation over carrying out practical constructive mathematics. In an effort to save constructive mathematics from the disdain afforded it by its association with intuitionism, Bishop set out to rebuild constructive analysis in a manner that captured both the criticisms of the constructivists and the essence of classical analysis. Thus Bishop published the \textit{Foundations of Constructive Analysis} in 1967, and, later, an updated volume with Douglas S. Bridges \cite{bishopBridges}, which we follow in this chapter with added commentary. It may please the reader to know that every numbered result in this section has been formally verified in this thesis's library \cite{libraryGit}.

\section{Sets, Operations, and Functions}
\label{setsopsfuncs}
A constructive theory of sets must require, of course, that its sets be ``constructible'' in some suitable sense. A set in Bishop's sense is thus defined by rules to construct an element of the set and an equivalence relation on the set that determines what it means for two of its elements to be equal. 

Thus Bishop's theory immediately distinguishes between intensional and extensional equality, where the extensional equality is given by a set's equipped equivalence relation. Bishop takes it a step further, defining an \textit{operation} from a set $A$ to a set $B$ by a ``finite routine'' intensionally mapping each element of $A$ to an element of $B$, and a \textit{function} from $A$ to $B$ by an operation from $A$ to $B$ that respects the extensional equality of $A$. An operation respects intension, and a function respects extension.

Those unfamiliar with constructive mathematics may be confused at why this distinction between operation and function is used. Why not take the quotient of a set over its equipped equivalence relation instead? Simply put, Bishop thinks that quotients would be pointless in his constructive setting \cite[p.~12]{bishopBridges}. Recall that sets are defined by rules to be satisfied to construct an element. An equivalence class $A$ (which is nonempty by definition) would thus be defined by \textit{explicitly providing} an element $x$ of $A$ and the equivalence relation. Hence, the boilerplate that comes with the equivalence class adds nothing, and we instead prove that operations like addition are functions in the sense we have defined. This may not be a satisfying conclusion, but there are other reasons for avoiding quotients and equivalence classes which we will see when discussing proof assistants in Chapter \ref{Agda}.

Finally, we note that the underlying logic of Bishop's system is essentially intuitionistic logic under the Brouwer-Heyting-Kolmogorov interpretation. We will now discuss some fundamental definitions and theorems regarding Bishop's real numbers, sequences, and series, taken from Bishop and Bridges \textit{Constructive Analysis} \cite{bishopBridges}.

\section{The Set of Real Numbers}
We will define the real numbers and their equality, and explore some of their properties. Note that we are using Bishop's conception of sets, operations, and functions, along with the BHK interpretation of logic.

\begin{definition}
\label{realDefinition}
A sequence $(x_n)$ of rational numbers is \textit{regular} if for all $m,n\in\mathbb{N}$ we have
\begin{equation}
\label{realRegularity}
|x_m - x_n| \leq \frac{1}{m} + \frac{1}{n}.
\end{equation}
A \textit{real number} is a regular sequence of rational numbers. Two real numbers $x\equiv(x_n)$ and $y\equiv(y_n)$ are $equal$, written $x = y$, if for all $n\in\mathbb{N}$ we have
\begin{equation}
\label{realEquality}
|x_n - y_n| \le \frac{2}{n}.
\end{equation}
\end{definition}

While reflexivity and symmetry are obvious, it is not as clear that equality of real numbers is transitive. It is a consequence of the following result.

\begin{proposition}
\label{equalityLemma}
The real numbers $x\equiv(x_n)$ and $y\equiv(y_n)$ are equal if and only if for each $j\in\mathbb{N}$ there is $N_j\in\mathbb{N}$ such that for all $n \ge N_j$ we have
\begin{equation}
\label{equalityLemmaHyp}
|x_n - y_n| \leq \frac{1}{j}.
\end{equation}
\begin{proof}
Suppose $x = y$. For each $j\in\mathbb{N}$, take $N_j \equiv 2j$. The desired result follows directly from (\ref{realEquality}).

Conversely, suppose that, for each $j\in\mathbb{N}$, there is $N_j\in\mathbb{N}$ such that for all $n \ge N_j$ we have (\ref{equalityLemmaHyp}). To show $x = y$, we must show that (\ref{realEquality}) holds. Consider an arbitrary $k\in\mathbb{N}$ and let $j \equiv 3k$. By assumption, there exists $N_j\in\mathbb{N}$ such that
\[
|x_{N_j} - y_{N_j}| \le \frac{1}{j} = \frac{1}{3k}.
\]
Without loss of generality, we can assume that $N_j \ge j$.
Using this and regularity of $(x_n)$ and $(y_n)$, we have
\begin{align*}
|x_n - y_n|
&= |x_n - y_n + (x_{N_j} - y_{N_j}) - (x_{N_j} - y_{N_j})|\\
&\le |x_n - x_{N_j}| + |x_{N_j} - y_{N_j}| + |y_n - y_{N_j}|\\
&\le (\frac{1}{n} + \frac{1}{N_j}) + \frac{1}{j} + (\frac{1}{n} + \frac{1}{N_j})\\
&\le \frac{1}{n} + \frac{1}{3k} + \frac{1}{3k} + \frac{1}{n} + \frac{1}{3k}\\
&= \frac{2}{n} + \frac{1}{k},
\end{align*}
so
\[
|x_n - y_n| \le \frac{2}{n} + \frac{1}{k}.
\]
Since $k\in\mathbb{N}$ was arbitrary, it follows that
\[
|x_n - y_n| \le \frac{2}{n},
\]
as desired.

\end{proof}
\end{proposition}

\begin{lemma}
Equality of real numbers is an equivalence relation.
\begin{proof}
Reflexivity and symmetry are obvious. For transitivity, let $x \equiv (x_n),y \equiv (y_n),z \equiv(z_n)$ be regular sequences, and suppose $x=y$ and $y=z$. Let $j\in\mathbb{N}$ be arbitrary and set $k \equiv 2j$. By (\ref{equalityLemma}), there exists $N_k\in\mathbb{N}$ such that for all $n \ge N_k$ we have
\[
|x_n - y_n| \le \frac{1}{k} = \frac{1}{2j}
\]
and $N_k'\in\mathbb{N}$ such that for all $n \ge N_k'$ we have
\[
|y_n - z_n| \le \frac{1}{k} = \frac{1}{2j}.
\]
Set $N_j \equiv \text{max}\{N_k, N_k'\}$. Then for all $n \ge N_j$ we have
\begin{align*}
|x_n - z_n| &\leq |x_n - y_n| + |y_n - z_n|\\
&\leq \frac{1}{k} + \frac{1}{k}\\
&= \frac{1}{2j} + \frac{1}{2j}\\
&=\frac{1}{j}.
\end{align*}
Since $j\in\mathbb{N}$ was arbitrary, it follows from (\ref{equalityLemma}) that $x = z$.
\end{proof}
\end{lemma}

From now on, we write $\mathbb{R}$ to refer to the set of real numbers, that is, the collection of regular sequences equipped with the equivalence relation defined in Definition \ref{realDefinition}.

To see what real number arithmetic looks like with regular sequences, we will define addition.

\begin{proposition}
Given real numbers $x\equiv (x_n)$ and $y\equiv (y_n)$, their sum
\[
(x_n) + (y_n) \equiv (x_{2n} + y_{2n})
\]
is a real number.
\begin{proof}
Let $m,n\in\mathbb{N}$. By regularity of $(x_n)$ and $(y_n)$, we have
\begin{align*}
|x_{2m} + y_{2m} - (x_{2n} + y_{2n})|
&\le |x_{2m} - x_{2n}| + |y_{2m} - y_{2n}|\\
&\le \frac{1}{2m} + \frac{1}{2n} + \frac{1}{2m} + \frac{1}{2n}\\
&= \frac{1}{m} + \frac{1}{n},
\end{align*}
so $(x_{2n} + y_{2n})$ is regular.
\end{proof}
\end{proposition}

Next, we define the order on $\mathbb{R}$.

\begin{definition}
Let $x,y\in\mathbb{R}$. We write $x<y$ if there exists $n\in\mathbb{N}$ such that
\[
y_n - x_n > \frac{1}{n}.
\]
We write $x\leq y$ if, for all $n\in\mathbb{N}$, we have
\[
y_n - x_n \geq -\frac{1}{n}.
\]
\end{definition}

Notice that we do not define $x\leq y$ to mean $x<y\lor x=y$. This is a double-edged sword. If we were to define $x\leq y$ in such a manner, then $\neg(x<y)$ would no longer imply $y\leq x$, since our constructive interpretation of logic would require us to prove either $y < x$ or $y = x$. This is not possible constructively. But the current definition avoids this problem. 

\begin{proposition}
Let $x,y\in\mathbb{R}$. If $\neg(x<y)$, then $x \geq y$.
\begin{proof}
Let $n\in\mathbb{N}$. Suppose $x_n - y_n < -\frac{1}{n}$. Then $y_n - x_n > \frac{1}{n}$, and so $x<y$, contradicting our assumption. Thus $x_n - y_n \ge -\frac{1}{n}$, giving $x \geq y$.
\end{proof}
\end{proposition}

Note that we freely used the fact that $\neg(p<q)$ implies $p \ge q$ for $p,q\in\mathbb{Q}$. This statement is easily proved constructively, but we needed to be careful with our definitions to prove it for $\mathbb{R}$. Next, it is easy to constructively prove that $\le$ is a total order on $\mathbb{Q}$, but it cannot be constructively shown that $\le$ is a total order on $\mathbb{R}$. Finally, equality is decidable on $\mathbb{Q}$, but not on $\mathbb{R}$.

Naturally, one may wonder what other subtleties are hidden in constructive mathematics. One point that is clear from our discussion of the differences between $\mathbb{Q}$ and $\mathbb{R}$ is that, for $x,y\in\mathbb{R}$, the statement $\neg (x=y)\implies x<y\lor x>y$ is not constructively valid. Thus for $x,y\in\mathbb{R}$ we write $x <> y$ if $x<y \lor y<x$. In place of the classical use of $x \neq y$ (i.e., $\neg(x = y)$), we tend to use $<>$, since that is usually what we want from $x \neq y$. Bishop takes it a step further, writing $x \neq y$ instead of $x <> y$ for $x<y \lor y<x$, but this is confusing by modern notational standards. We will encounter similar strangeness in the treatment of sequences and series.

Despite these oddities, it is worth noting that Bishop's continuum is consistent with the classical continuum, in that it is uncountable (Cantor's technique is constructive), and both the Archimedean property and the density of $\mathbb{Q}$ in $\mathbb{R}$ are constructively valid (we have verified their proofs in the Agda proof assistant, along with all propositions, lemmas, and theorems in this chapter). But is Cauchy completeness constructively valid?

\section{Sequences}
Any constructive treatment of sequences that cannot prove Cauchy completeness of $\mathbb{R}$ is bound to be rejected in favour of the classical treatment. Classically, we often prove completeness by ultimately relying on the least upper bound property. But for the supremum of a nonempty subset $S$ of $\mathbb{R}$ that is bounded above to exist constructively, we would need to provide an algorithm to construct it, a tall order given that we know essentially nothing about the elements of $S$. Indeed, the least upper bound property is constructively invalid. Fortunately, there is a valid constructive proof of completeness in Bishop's system, though with some peculiarities we will later note.

\begin{definition}
\label{cauchySequenceDefinition}
A sequence $(x_n)$ of real numbers is a \textit{Cauchy sequence} if for each $k\in\mathbb{N}$ there is $N_k\in\mathbb{N}$ such that for all $m,n\ge N_k$ we have
\begin{equation}
\label{cauchySequenceInequality}
|x_m - x_n| \leq \frac{1}{k}.
\end{equation}
\end{definition}

\begin{definition}
\label{convergenceDefinition}
A sequence $(x_n)$ of real numbers \textit{converges} to a real number $l$ if for each $k\in\mathbb{N}$ there is $N_k\in\mathbb{N}$ such that $n\ge N_k$
\begin{equation}
\label{convergenceInequality}
|x_n - l| \le \frac{1}{k}.
\end{equation}
We say that $(x_n)$ is \textit{convergent} if there is a real number $l$ such that $(x_n)$ converges to $l$.
\end{definition}

In Bishop's proof of completeness of $\mathbb{R}$ we will need to know, for a real number $x\equiv (x_n)$, how accurately each $x_n$ approximates $x$. We state the result without proof, for which the reader may refer to Lemma 2.14 in \cite[p.~25]{bishopBridges} or the \verb|lemma-2-14| function in the \verb|RealProperties.agda| file of \cite{libraryGit}.

\begin{proposition}
\label{rationalApproximation}
For each real number $x\equiv (x_n)$, we have
\[
|x - x_n| \le \frac{1}{n}
\]
for all $n\in\mathbb{N}$.
\end{proposition}

We now present Bishop's proof of completeness with clarification added where it is helpful.

\begin{theorem}
\label{cauchyCompletenessTheorem}
A sequence $(x_n)$ of real numbers converges if and only if it is a Cauchy sequence.
\begin{proof}
Suppose $(x_n)$ converges to some $l\in\mathbb{R}$. Let $k\in\mathbb{N}$. For $(x_n)$ to be a Cauchy sequence, we must show that there is $N_k\in\mathbb{N}$ such that for all $m,n\ge N_k$ we have
\[
|x_m - x_n| \le \frac{1}{k}.
\]
By definition of convergence, there exists $M_{2k}\in\mathbb{N}$ such that for all $m,n\ge M_{2k}$ we have
\[
|x_n - l| \le \frac{1}{2k}.
\]
Take $N_k \equiv M_{2k}$. Then for all $m,n \ge N_k$ we have
\begin{align*}
|x_m - x_n|
&\le |x_m - l| + |x_n - l|\\
&\le \frac{1}{2k} + \frac{1}{2k}
= \frac{1}{k},
\end{align*}
and we are done.

Conversely, suppose $(x_n)$ is a Cauchy sequence. We will first construct a real number $y$, and then show that $(x_n)$ converges to $y$. We define $y \equiv (y_k)$ as follows. Consider some $k\in\mathbb{N}$. By Definition \ref{cauchySequenceDefinition}, there exists $N_{2k}\in\mathbb{N}$ such that for all $i,j \ge N_{2k}$, we have
\[
|x_i - x_j| \le \frac{1}{2k}. \tag{$\star$}
\]
Let $M_k \equiv \text{max}\{3k,N_{2k}\}$. Consider the real number $x_{M_k}$ in the sequence $(x_n)$. We define $y_k \equiv x_{M_k,2k}$; that is, $y_k$ is the $(2k)$\textsuperscript{th} element of the regular sequence defining $x_{M_k}$. Since $k\in\mathbb{N}$ was arbitrary, we define the entire sequence $y \equiv (y_k)$ through this process.

To show that $y$ is a real number, we must show that $(y_k)$ is a regular sequence. To this end, let $m,n\in\mathbb{N}$. First, note that we get $M_m, M_n \ge \text{min}\{N_{2m},N_{2n}\}$ by definition of $M_m$ and $M_n$. From $(\star)$, it follows that
\begin{align*}
|x_{M_m} - x_{M_n}|
&\le \text{max}\{\frac{1}{2m},\frac{1}{2n}\}\\
&\le \frac{1}{2m} + \frac{1}{2n}. \tag{$\star \star$}
\end{align*}
By Proposition \ref{rationalApproximation} and $(\star \star)$, we have
\begin{align*}
|y_m - y_n| &= |x_{M_m,2m} - x_{M_n,2n}|\\
&\le |x_{M_m,2m} - x_{M_m}| + |x_{M_m} - x_{M_n}| + |x_{M_n} - x_{M_n,2n}|\\
&\le \frac{1}{2m} + (\frac{1}{2m} + \frac{1}{2n}) + \frac{1}{2n}\\
&= \frac{1}{m} + \frac{1}{n}.
\end{align*}
Since $m$ and $n$ were arbitrary, it follows that $(y_k)$ is regular. Thus $y$ is a real number.

It remains to show that $(x_n)$ converges to $y$. To this end, let $k\in\mathbb{N}$ be arbitrary. It suffices to show that for all $n\ge M_k$ we have
\[
|y - x_n| \le \frac{1}{k}.
\]
Taking $n \ge M_k \equiv \text{max}\{3k, N_{2k}\}$, it follows that $n \ge 3k$, $M_n \ge N_{2k}$, and $n \ge N_{2k}$. Since $M_n, n \ge N_{2k}$, we can apply $(\star)$. Combined with Proposition \ref{rationalApproximation}, we have
\begin{align*}
|y - x_n| 
&\le |y - y_n| + |y_n - x_{M_n}| + |x_{M_n} - x_n|\\
&= |y - y_n| + |x_{M_n,2n} - x_{M_n}| + |x_{M_n} - x_n|\\
&\le \frac{1}{n} + \frac{1}{2n} + \frac{1}{2k}\\
&\le \frac{1}{3k} + \frac{1}{6k} + \frac{1}{2k}\\
&= \frac{1}{k},
\end{align*}
as desired.
\end{proof}
\end{theorem}

Before discussing the peculiarities of Bishop's notions of Cauchyness and convergence, we must discuss the status of the axiom of choice and its weaker counterpart, the axiom of countable choice. We thus repeat and extend the comments of Douglas Bridges in \cite[p.~13]{bishopBridges}.

For our purposes, the axiom of choice (AoC) states that, for sets $X$ and $Y$ and a relation $R \subseteq X\times Y$, if for each $x\in X$ there exists $y\in Y$ such that $xRy$ holds, then there is a function $f : X \to Y$ such that for each $x\in X$ we have $xRf(x)$. The axiom of countable choice (AoCC) is obtained from AoC by replacing $X$ with $\mathbb{N}$.

By Diaconescu's theorem, AoC implies the law of the excluded middle (LEM). Thus, AoC cannot be constructively valid. However, Bishop himself seemingly states otherwise, stating that ``A choice function exists in constructive mathematics, because a choice is \textit{implied by the very meaning of existence}'' \cite[p.~13]{bishopBridges}. Indeed, AoC is, in a weaker sense, true in Bishop's system, as is AoCC.

The key lies in Bishop's distinction between ``function'' and ``operation'', and, moreover, in what is meant by ``there exists''. In the classical sense of ``function'', the function $f : X \to Y$ in AoC is understood to respect the extensional equality on $X$, and thus the classical notion of ``function'' is essentially equivalent to Bishop's notion. If we do not require $f$ to be a function in Bishop's sense, and only an operation, then Diaconescu's theorem no longer applies.

This weak AoC (wAoC), where $f$ is merely an operation, is actually provable in Bishop's system. Recall that ``there exists an $x$ such that $P(x)$'' means that we have a pair of some $x$ and a proof of $P(x)$. Thus for sets $X,Y$ and a relation $R \subseteq X\times Y$, the statement ``for each $x\in X$ there is $y\in Y$ such that $xRy$'' literally means that there is an operation $f : X \to Y$ such that we have $xRf(x)$ for each $x\in X$, and so wAoC is one of the most trivial theorems in Bishop's system. AoCC is then a corollary of wAoC since the notion of equality of two natural numbers is intensional equality.

It turns out that AoCC is critical for $\mathbb{R}$ to be Cauchy complete constructively. Again analyzing ``there exists'', Bishop's definition of ``$(x_n)$ is a Cauchy sequence'' boils down to the following (the corresponding interpretation of convergence is obvious).

\begin{definition}
\label{analyzedCauchyness}
A sequence of reals $(x_n)$ is a \textit{Cauchy sequence} if there exists a function $N : \mathbb{N} \to \mathbb{N}$ such that, for all $k\in\mathbb{N}$ and $m,n\ge N(k)$ we have
\[
|x_m - x_n| \le \frac{1}{k}.
\]
\end{definition}

Robert Lubarsky \cite{lubarsky} showed that if AoCC is invalid in some constructive settings and if there is not necessarily a function $N : \mathbb{N} \to \mathbb{N}$ as in Definition \ref{analyzedCauchyness} (i.e., if we cannot interpret Definition \ref{cauchySequenceDefinition} as Definition \ref{analyzedCauchyness}), it may be impossible to show that the reals are complete. Thus constructivizations of analysis weaker than Bishop's system tend to be untenable.

\section{Conclusion}
While Bishop's system comes with oddities like the fact that $\le$ is not a total order on $\mathbb{R}$, we see that constructive analysis is not as untenable as Hilbert thought. Not only does Bishop's system avoid the philosophical baggage of Brouwer's intuitionism, Bishop's continuum is quite similar to the classical continuum, satisfying uncountability, density of $\mathbb{Q}$, the Archimedean property, and completeness. And most importantly, Bishop's treatment of analysis is in-depth, covering calculus, complex analysis, and beyond. But these topics and their fascinating subtleties are far outside the scope of this thesis. 

Before continuing on to Agda, the reader is reminded that every numbered result in this chapter has been formally verified in my constructive analysis library \cite{libraryGit}.

\chapter{Agda}\label{Agda}
\section{Introduction}
A \textit{proof assistant} is a software tool for writing mathematical proofs in such a manner that a computer can verify their validity. Modern proof assistants often use a \textit{type theory} powerful enough to express interesting mathematics via the \textit{Curry-Howard correspondence}. Type theory is like set theory, but it is more convenient for automation. By \textit{type}, we mean the types of a programming language, like ``char'' or ``int''. Mathematical propositions are expressed as types, and proofs are programs that construct a term with the proposition as its type. For example, the proposition $P \implies Q$ is represented by the function type $P \to Q$. A proof of $P \implies Q$ is then a function mapping proofs of $P$ (i.e., terms of type $P$) to proofs of $Q$ (i.e., terms of type $Q$).
In sum, propositions are types and proofs are programs. This is precisely the Curry-Howard correspondence. Note that it is strikingly similar to the Brouwer-Heyting-Kolmogorov interpretation of logic.

Such proof assistants usually include a programming language in which we can exploit features of computation, such as recursion and reflection, to automate portions of proofs. Three of the most prominent proof assistants are Lean, Coq, and Agda \cite{lean,coq,agda}. We will focus on Agda, a dependently typed functional programming language based on a type theory similar to Martin-L\"{o}f's intensional type theory. We will discuss Agda's basic use, features, and problems that a typical mathematician user may encounter. Some familiarity with the lambda calculus is assumed in this chapter, for which the reader may refer to \cite{nederpeltGeuvers}.

\section{Definitions and Proofs in Agda}\label{section:defProofsAgda}
We will give a brief introduction to Agda through some examples. Agda provides plenty of helpful features for the user to define their objects and make their proofs at least somewhat readable. We begin by defining the Peano numbers.

\begin{definition}
Peano axiomatized the set $\mathbb{N}$ of natural numbers as follows.
\begin{itemize}
	\item $0$ is a natural number.
	\item Whenever $n$ is a natural number, so is $suc(n)$.
	\item $0 \neq suc(n)$ for all $n\in\mathbb{N}$.
	\item If $suc(m) = suc(n)$, then $m = n$.
	\item For any set $S$, if $0\in S$ and if for all $n\in\mathbb{N} \cap S$ we have $suc(n)\in S$, then $\mathbb{N} \subseteq S$.
\end{itemize}
\end{definition}

In Agda, we can concisely define the natural numbers as a data type as follows.

\begin{center}
\begin{BVerbatim}
data ℕ : Set where 
  zero : ℕ
  suc  : ℕ → ℕ
\end{BVerbatim}
\end{center}

Note that \verb|ℕ| has type \verb|Set|, indicating that the natural numbers are a set. The elements past the \verb|where| keyword are the type constructors of \verb|ℕ|, indicating that \verb|zero| is a natural number and that \verb|suc : ℕ → ℕ| is a function producing a natural number \verb|suc n : ℕ| given \verb|n : ℕ|. The \verb|data| keyword indicates that the definition is inductive, meaning that the only elements of \verb|ℕ| are those built with the specified type constructors.

Typically, addition on $\mathbb{N}$ is defined recursively on natural numbers $m,n\in\mathbb{N}$ as follows.
\begin{align*}
0 + n &= n\text{, and}\\
suc(m) + n &= suc(m + n).
\end{align*}
Since \verb|ℕ| was defined inductively, Agda allows us to recursively define addition in precisely the same way by pattern matching on the type constructors of \verb|ℕ|. Note that the definition is \textit{curried}, meaning that instead of defining a function $A \times B \to C$, we define a function $A \to (B \to C)$. Curried definitions are the preferred style in Agda.

\begin{center}
\begin{BVerbatim}
_+_ : ℕ → ℕ → ℕ
zero + n  = n
suc m + n = suc (m + n)
\end{BVerbatim}
\end{center}

To illustrate a proof, let us prove the commutativity of addition. For the sake of comparison with Agda, we first write the proof informally.

\begin{proposition}
For each $m,n\in\mathbb{N}$, we have $m + n = n + m$.
\begin{proof}
We proceed by induction on $m$ and $n$. We consider several cases.
\begin{enumerate}[label=\textbf{Case \arabic*.}, wide=0pt]
\item Suppose $m = n = 0$. Then 0 + 0 = 0 + 0 holds by reflexivity.

\item Suppose $m = 0$. Our induction hypothesis is that $0 + n = n + 0$. We must show that $0 + suc(n) = suc(n) + 0$. By definition of $+$ and $suc$, this is equivalent to $suc(0 + n) = suc(n + 0)$, which follows from the induction hypothesis and the fact that $suc$ is a function. The case where $n = 0$ with induction hypothesis $m + 0 = 0 + m$ is similar. 

\item Our induction hypotheses are the following:

\begin{enumerate}[label=(\roman*)]
	\item $m + suc(n) = suc(n) + m$,
	\item $m + n = n + m$, and
	\item $suc(m) + n = n + suc(m)$.
\end{enumerate}

We must show that $suc(m) + suc(n) = suc(n) + suc(m)$. We have
\begin{align*}
suc(m) + suc(n) 
&= suc(m + suc(n)) \qquad\text{by definition,}\\
&= suc(suc(n) + m) \qquad\text{since suc is a function, by hypothesis (i),}\\
&= suc(suc(n + m)) \qquad\text{by definition,}\\
&= suc(suc(m + n)) \qquad\text{since suc is a function, by hypothesis (ii),}\\
&= suc(suc(m) + n) \qquad\text{by definition,}\\
&= suc(n + suc(m)) \qquad\text{since suc is a function, by hypothesis (iii),}\\
&= suc(n) + suc(m) \qquad\text{by definition,}
\end{align*}
\end{enumerate}
and we are done.
\end{proof}
\end{proposition}

The proof of commutativity of addition is compactly translated into Agda. By the Curry-Howard correspondence, the statement ``$\forall m,n\in\mathbb{N}. m+n = n+m$'' is encoded in Agda by the following function and type.
\begin{center}
\begin{BVerbatim}
+-comm : (m n : ℕ) → m + n ≡ n + m
\end{BVerbatim}
\end{center}
First, note that we represent the universal quantification over $m$ and $n$ by a function taking inputs \verb|m| and \verb|n|. The commutativity proposition regarding these inputs is the codomain of this function. Note that $\equiv$ is used in place of $=$, since the symbol $=$ is used for definitions in Agda. To prove commutativity, we recursively define \verb|+-comm| on the type constructors \verb|zero| and \verb|suc|.

First, we simply have to prove $0 + 0 = 0 + 0$, which is true by reflexivity. The \verb|refl| term proves reflexivity for any given term, and so we use it here to prove \verb|zero + zero ≡ zero + zero| as follows.

\begin{center}
\begin{BVerbatim}
+-comm zero zero = refl
\end{BVerbatim}
\end{center}

Next, we prove that $0 + suc(n) = suc(n) + 0$ by defining \verb|+-comm| on the values \verb|zero| and \verb|suc n|. Note that the induction hypothesis is condensed into a recursive call to the function \verb|+-comm|.

\begin{center}
\begin{BVerbatim}
+-comm zero (suc n) = cong suc (+-comm 0 n)
\end{BVerbatim}
\end{center}

Agda requires a term of type \verb|zero + suc n ≡ suc n + zero|. Since the term \verb|zero + suc n| is definitionally equivalent to \verb|suc n| and \verb|suc n + zero| is definitionally equivalent to \verb|suc (n + zero)|, Agda replaces these terms by their definitions automatically and requires, instead, a proof of \verb|suc n ≡ suc (n + zero)|. In our informal proof, we used the fact that $suc$ is a function. We do so in our Agda proof as well by using the \verb|cong| function. We supply it our function, \verb|suc|, and we then need to provide a term (or proof) of type  \verb|n ≡ n + zero| as the second argument to \verb|cong|. We provide the second argument via recursion, imitating the induction step in our informal proof. The proof of $suc(m) + 0 = 0 + suc(m)$ is similar.

Finally, we show that $suc(m) + suc(n) = suc(n) + suc(m)$. To this end, we recursively define \verb|+-comm| on the arguments \verb|suc m| and \verb|suc n|.
\newpage
\begin{center}
\begin{BVerbatim}
+-comm (suc m) (suc n) = begin
  suc m + suc n     ≡⟨ refl ⟩
  suc (m + suc n)   ≡⟨ cong suc (+-comm m (suc n)) ⟩
  suc (suc n + m)   ≡⟨ refl ⟩
  suc (suc (n + m)) ≡⟨ cong suc (cong suc
                                (+-comm n m)) ⟩
  suc (suc (m + n)) ≡⟨ refl ⟩
  suc (suc m + n)   ≡⟨ cong suc (+-comm (suc m) n) ⟩
  suc (n + suc m)   ≡⟨ refl ⟩
  suc n + suc m      ∎
\end{BVerbatim}
\end{center}

Thankfully, someone built an equational reasoning package in Agda. It provides a convenient and readable syntax that looks much like our informal proof. Such a proof starts with the \verb|begin| keyword and ends with the QED symbol \verb|∎|, requiring various steps in between. The most notable difference is that instead of writing the equation first and the reason after, we write one side of the equation, the reason, then the other side of the equation. For instance, the step that \verb|suc m + suc n ≡ suc (m + suc n)| is proved via \verb|refl|, and is read, in the Agda proof, like ``$suc(m) + suc(n)$, by reflexivity, is equal to $suc (m + suc(n))$''.

The reader may wonder if it is necessary to apply the reflexivity steps in our equational proof. The answer is no: They may be removed, since Agda computes all terms to their definitions automatically. Our final definition of \verb|+-comm| thus looks like the following.
\newpage
\begin{center}
\begin{BVerbatim}
+-comm : (m n : ℕ) → m + n ≡ n + m
+-comm zero zero       = refl
+-comm zero (suc n)    = cong suc (+-comm 0 n)
+-comm (suc m) zero    = cong suc (+-comm m 0)
+-comm (suc m) (suc n) = begin
  suc (m + suc n)   ≡⟨ cong suc (+-comm m (suc n)) ⟩
  suc (suc (n + m)) ≡⟨ cong suc (cong suc
                                (+-comm n m)) ⟩
  suc (suc m + n)   ≡⟨ cong suc (+-comm (suc m) n) ⟩
  suc n + suc m      ∎
  where open ≡-Reasoning
\end{BVerbatim}
\end{center}

Lastly, we note the \verb|where open ≡-Reasoning| string. The \verb|where| clause opens a new block of code local to the function being defined, which retains access to the variables and terms being used by the function (such as its parameters). Inside a \verb|where| clause, we can define functions, prove lemmas, or, in this case, open the \verb|≡-Reasoning| package that lets us use the equational reasoning format shown.

The reader, perhaps noticing how often we state that ``Agda reduces these terms automatically, so we prove [this other equation] instead'', may wonder if the user has to mentally track what is being proved, and how horrific this must be in cases where the proofs are much more complex! Perhaps Agda's greatest feature is its interactive mode for the Emacs environment, built to solve these kinds of inconveniences. 
Consider the proof of $0 + suc(n) = suc(n) + 0$. When writing the corresponding definition of \verb|+-comm|, the user may replace the right side of the definition by a hole, represented by \verb|{ }|. 

\begin{center}
\begin{BVerbatim}
+-comm zero (suc n) = { }
\end{BVerbatim}
\end{center}

If the user clicks the hole, they have at least three options, depending on the keyboard command they enter. One, they could view the current \textit{goal} in an Emacs subwindow, which is what must be proved (the subwindow also shows the variables and hypotheses the user has access to for the function they are defining). Two, they could enter a term, and by entering different commands Agda will display its type and compute its value. And three, once a term is entered and satisfies the goal, the hole may be \textit{refined}, replacing the hole by the term given and completing the user's proof or definition. 

There are a number of other features regarding Agda's interactive Emacs mode, but holes are perhaps the most convenient. Without them, the user would have to keep track of potentially thousands of goals and potentially massive types mentally or on paper! 

The reader may wonder, ``What else can Agda automate? Surely it is not just definitions!'' Indeed, Agda can automate entire proofs in some cases. For instance, the Agda standard library contains a ring solver which, given a (true) ring equation, provides a proof. The precise function of the ring solver is beyond the scope of this thesis, but it essentially takes both sides of a polynomial equation, reduces them to a form called a \textit{Horner normal form}, and checks that the forms of both sides are equal by reflexivity. Of course, the ring solver is itself programmed in Agda, so not only does Agda provide this excellent feature, it also proves its validity.

So far, we have seen basic data types, inductive proofs, the package system, and the immense amount of automation in the forms of replacing terms by their definitions, ring solvers, and holes. We have exploited the computational features of the type system, recursion, and an IDE-like environment. The vast power of Agda does not end here. For instance, the cubical Agda mode enables the use of homotopy type theory. To learn more about what Agda has to offer, please visit the course at \cite{peterCourse} and Agda's documentation at \cite{agda}.

\section{Problems with Agda}
\subsection{Canonicity}\label{canonicity}

We say that our type theory has \textit{canonicity} if every term can be reduced to some canonical form (e.g., via a series of $\beta$-reductions applied to the term). Agda, for instance, has canonicity. 

Canonical forms are extremely useful in proof assistants. Reducing two terms to their canonical forms, for instance, can help us decide if they are equal. Consider the proof that $0 + 1$ equals $1 + 0$. Because Agda computes both $0 + 1$ and $1 + 0$ to their canonical form, $1$, we can easily state that the two are equal by reflexivity. The ring solver also relies, to some extent, on canonicity, as it computes the two sides of an equation into their canonical forms and checks for reflexivity to decide if they are equal.

Losing canonicity can be problematic. However, it can be equally, if not more problematic to preserve canonicity entirely. Let us define a function $f : \mathbb{R} \to \mathbb{N}$ such that
\[
f(x) =
\begin{cases}
0 & \text{if } x = 0,\\
1 & \text{if } x \neq 0.
\end{cases}
\] 

Since, in general, it is impossible to decide whether or not a real number $x$ equals $0$, $f(x)$ may never be computed. Thus an equation about natural numbers involving $f(x)$ may not reduce to a canonical form, and we may be unable to use reflexivity in a proof involving $f(x)$. If we decide to preserve canonicity and forbid such definitions, we naturally must abandon axioms like the law of the excluded middle (LEM). The addition of axioms, in general, will extinguish canonicity. Preservation of canonicity at all costs is thus an extreme form of constructivism.

However, the extent to which canonicity is an issue is dependent on the mathematics being done and on the programming. Mathematics that significantly relies on axioms and LEM will naturally face canonicity issues in a proof assistant. It is the point of Bishop's constructive analysis that analysts are not as bound to LEM as we may believe. For instance, we usually define the absolute value $|x|$ of a real number $x$ by
\[
|x| =
\begin{cases}
x & \text{if } x \geq 0,\\
-x & \text{if } x \ngeq 0.
\end{cases}
\]
If we then wish to prove $|x| < y$ by proving that $x < y$ and $-x < y$, we need to invoke LEM to decide whether $x \geq 0$ or $x \ngeq 0$. An alternative (and classically equivalent) constructive approach exists. We can easily define the maximum of real numbers $x$ and $y$ constructively by
\[
(\text{max}\{x,y\})_n = \text{max}\{x_n,y_n\}
\]
and thus define
\[
|x| = \text{max}\{x,-x\}.
\]
Using properties of regular sequences, we can show that if $x < y$ and $-x < y$, then $|x| < y$. This time, LEM is unneeded. Bishop details many more examples where LEM is unnecessary. Hence, canonicity issues are, to a possibly great extent, dependent upon the implementation of our mathematics. 

Sometimes, it is useful to forego canonicity for reasons other than the kind of mathematics being done or the mathematician's technique. For the sake of performance, we often want our terms to \textit{not} compute to a canonical form! For example, consider the Archimedean property. If all we care about is that some natural $N$ greater than a fixed real number $x$ exists, and not explicitly what $N$ is, why waste time computing $N$ when we can extract this information from the statement of the Archimedean property itself? There are practical cases where computing $N$ takes days, even in our own constructive analysis library (the Archimedean property is a nuisance for performance in particular). Agda thus allows us to mark functions as \verb|abstract|, meaning they will not be computed, but their types can still be used and manipulated as desired.

\subsection{Setoids and Quotient Sets}

In Chapter \ref{setsopsfuncs}, we noted that, in some sense, quotient sets are constructively redundant. But a user doing classical mathematics in Agda would probably like to use quotient sets. Unfortunately, adding quotient types to Agda can break canonicity. In modern proof assistants that focus more on classical mathematics, like Lean, types representing quotient sets, known as \textit{quotient types}, are axiomatized, and canonicity is lost. Thus Agda users (and many proof assistant users in general) who wish to emulate quotients without losing canonicity typically resort to using Bishop's conception of a set, described in Chapter \ref{setsopsfuncs}. For the sake of clarity, Bishop's sets are widely referred to as \textit{setoids}. Thus, a setoid is (classically) a set equipped with an equivalence relation on its elements.

But setoids are impractical, to the extent that the proof assistant community coined the term ``Setoid Hell'' to refer to their use. For a given set $S$ with an equivalence relation $\sim$, the quotient of $S$ by $\sim$ enables us to discuss equality of elements of $S$ without referring to $\sim$, which is hidden behind equivalence classes. But without quotients, we \textit{need} the equivalence relation $\sim$ to discuss any meaningful notion of equality on $S$. Instead of one notion of equality, we have one for each set, which is obviously quite cumbersome.

For Agda users, there is a third alternative, in which quotients can be defined without losing canonicity: cubical Agda. Homotopy type theory (HoTT), and the ``univalent foundations for mathematics'' in general, is an alternative foundation of mathematics formulated chiefly by the Fields medalist Vladimir Voevodsky. A description of HoTT is far beyond the scope of this thesis. Readers interested in HoTT should refer to \cite{hottBook}. Cubical Agda essentially uses HoTT, unlike normal Agda which uses a version of Martin-L\"{o}f's intensional type theory.

The version of HoTT that comes with cubical Agda can be used to represent quotients without breaking canonicity. However, swapping foundations may seem heavy-duty. Moreover, the status of HoTT as a foundation for mathematics is still debated. But at present, cubical Agda is the only way to obtain quotients while preserving canonicity.

\subsection{Conclusion}

The issues of canonicity and quotient types are likely to be far removed from many mathematicians' research. But anyone using Agda as a proof assistant is bound to encounter these problems eventually, on top of any issues with performance that naturally come with Agda being a programming language. Theoretical problems made practical like these show us that proof assistants are not as ready to replace {\LaTeX} as a proof assistant expert would like. But it is a testament to the usefulness and potential of proof assistants that, despite these issues, their communities and their libraries are growing quickly. For instance, the Lean proof assistant began at Microsoft Research, and the Agda standard library has a growing category theory library \cite{agdaCategories}. Regardless of whether or not proof assistants replace \LaTeX , their use in proving otherwise unwieldy results like the four colour theorem \cite{gonthier} shows that using a proof assistant is a useful skill, one that often lies at the intersection of programming and mathematics.

\chapter{Constructive Analysis in Agda}\label{constructiveanalysisinagda}

We will now discuss our constructive analysis library in Agda. From Bishop and Bridges' \textit{Constructive Analysis}, we have completed sections 2 and 3 of Chapter 2 (save a couple of minor exceptions), thus covering the real numbers and their basic properties (including uncountability), and results regarding sequences and series (such as Cauchy completeness and tests for convergence). Some work on the following section, covering continuity, was also done. Finally, a plethora of lemmas, corollaries, theorems, and propositions not present in \textit{Constructive Analysis} were proved. Many were required to prove results from \textit{Constructive Analysis} but were not given in the book, while others helped automate portions of proofs or improved performance. The library can be found at \cite{libraryGit}. The exact list of files is as follows.

\begin{table}[tbh]
\centering
\begin{tabular}{|p{4cm}|p{10cm}|}
	\hline
	File Name & File Description
	\\
	\hline
	ExtraProperties.agda & A collection of extra properties about naturals, integers, and rationals that were useful in the rest of the library. \\
	\hline
	NonReflective.agda & The base ring solver file.\\
	\hline
	NonReflectiveQ.agda & The ring solver instantiated for the rational numbers. \\
	\hline
	NonReflectiveZ.agda & The ring solver instantiated for the integers. \\
	\hline
	Real.agda & Definition of the real numbers, arithmetic operations, and ordering. \\
	\hline
	RealProperties.agda & Properties of the real numbers, like arithmetic properties and the density of the rationals. \\
	\hline
	Inverse.agda & The definition of the multiplicative inverse and a number of its properties.\\
	\hline
	Uncountability.agda & The proof that the continuum is uncountable.\\
	\hline
	Sequence.agda & Definitions and properties regarding sequences and series, including results like the algebraic limit theorem and Cauchy completeness.\\
	\hline
	MetricBase.agda & Some work on defining metric spaces. Incomplete.\\
	\hline
	Continuity.agda & Some work on the definitions and properties of continuity. Incomplete.\\
	\hline
\end{tabular}
\caption{Agda files in the library.}
\label{tab:AgdaFiles}
\end{table}

\section{Why Constructive Analysis?}

Readers with little or no experience in Agda may wonder why someone would formalize results from constructive analysis. There are two distinct questions: One, why analysis? And two, why \textit{constructive} analysis?

I selected analysis because there is very little analysis in Agda. Agda users tend to be varying kinds of computer scientists, algebraists, and programmers. The mathematics they usually focus on, like category theory, requires little or no analysis. Hence, analysis is ignored.

I chose Bishop's constructive blend of analysis mainly for convenience. Recall the problems with axioms and canonicity in Agda. Classical analysis is highly reliant on non-constructive axioms. Thus classical analysis does not fit into the highly constructive setting of Agda as neatly as constructive analysis. If constructive analysis cannot be formalized in Agda, there is no hope for classical analysis!

\section{What Libraries Do We Need?}

Considering that Bishop's reals are defined as regular sequences of rational numbers, we obviously need integers and rational numbers. On top of these, we need some logic and basic automation tools for simple proofs. Conveniently, everything we needed was already present in the Agda standard library.

While the Agda standard library, at the time of writing, lacks analysis (and real numbers in general), it is rich in logic, integers, rational numbers, and various tools to automate the proofs of statements about each of these topics. The Brouwer-Heyting-Kolmogorov interpretation is native to Agda, and many other tools of logic are implemented by fellow Agda users in its standard library. The basic theory of functions and relations is spelled out richly. The integers are implemented rather nicely as an extension of the natural numbers (as previously defined) and their most important basic properties are proved and ready for use. Of course, as already mentioned in Section \ref{section:defProofsAgda}, the ring solver is available for automating the proof of valid ring equations.

On the other hand, there are two versions of the rational numbers in the Agda standard library. The \textit{normalized} rationals require that the numerator and denominator be coprime, while the \textit{unnormalized} rationals do not. Requiring that the numerator and denominator be coprime can place a large proof burden on the user. It is rare in informal mathematics, let alone formalized mathematics, to write a fraction with its numerator and denominator coprime. The normalized rationals thus include a function to automatically reduce a fraction to such a form. But this process can be computationally expensive. The normalized rationals are thus inefficient for our purposes. Hence, we use the unnormalized rationals.

With all of this, along with all of the Agda standard library's syntactic sugar, it was easy to start working on Bishop's constructive analysis right away.

\section{The Real Numbers and Software Engineering}

We now turn to the definition of the real numbers in Agda. In particular, we will discuss some software engineering used to improve their usability and the library's overall performance.

We note that the definition of a Bishop real number and equality on Bishop reals was first formalized in Agda by Martin Lundfall \cite{lundfall}. However, Lundfall's library only proved some very basic results, up to the fact that the Bishop's definition of real number equality is an equivalence relation. Our library, as previously mentioned, is much more extensive. We thus compare and contrast the engineering behind my implementation of Bishop's reals and Lundfall's implementation.

Recall that a real number, in Bishop's sense, is a pair consisting of a sequence of rationals and a proof that the sequence is regular. To represent this pairing in Agda, we use a \verb|record| type.

\begin{center}
\begin{BVerbatim}
record ℝ : Set where
  constructor mkℝ
  field
    seq : ℕ → ℚᵘ
    reg : (m n : ℕ) {m≢0 : m ≢0} {n≢0 : n ≢0} →
          ℚ.∣ seq m ℚ.- seq n ∣ ℚ.≤
          (+ 1 / m) {m≢0} ℚ.+ (+ 1 / n) {n≢0}
\end{BVerbatim}
\end{center}

Note that this defines a set called \verb|ℝ|. Real numbers have \textit{fields} (like how a Java object has field variables) called \verb|seq| and \verb|reg|, which represent a sequence of rationals and a proof that the sequence is regular, respectively. For convenience, a constructor \verb|mkℝ| is provided, so that a real number with sequence \verb|xs : ℕ → ℚᵘ| and proof of regularity \verb|P| may be constructed by writing \verb|mkℝ xs P|.

We note that \verb|ℚᵘ| represents the type of unnormalized rational numbers, as previously defined. To use, for instance, addition of rational numbers in the type of \verb|reg|, we write \verb|ℚ.+|. The \verb|ℚ.| represents that we are using rational number addition, not real number addition. Unfortunately, we could not overload the addition operator to write just \verb|+| for each kind of addition. The same goes for the other arithmetic operators.

We note a subtlety in the type corresponding to regularity. Rational numbers require denominators to be nonzero, so for each denominator, we must provide a proof that it is nonzero. Thus for the natural number \verb|m| we must provide a proof \verb|m≢0 : m ≢0| that it is not nonzero. The curly braces in \verb|{m≢0 : m ≢0}| represent that the proof that \verb|m| is nonzero is \textit{implicit}. Agda will attempt to automatically prove implicit parameters. Indeed, since the natural numbers are recursively defined and a natural number is nonzero when it is of the form \verb|suc n|, Agda will immediately recognize any natural number written in this form as nonzero. Thus, Agda does not require the user to provide this proof when writing \verb|reg| on natural numbers of the form \verb|suc n|. This also means that when writing a fraction \verb|+ 1 / m| as above, we need not supply the proof that \verb|m| is nonzero if it is written in the form \verb|suc n|. We only needed to supply the proof that \verb|m| was nonzero in the regularity type because \verb|m| does not compute to a form \verb|suc n| in the type given.

The regularity type in Lundfall's definition did not require proofs that the naturals \verb|m| and \verb|n| are nonzero. Lundfall's solution was to write \verb|suc m| and \verb|suc n| in the denominators instead of \verb|m| and \verb|n| \cite[p.~10]{lundfall}. As a result, Lundfall's definition looks somewhat cleaner.

Initially, we used the same strategy as Lundfall. But we found that given natural numbers of the form \verb|k| that were not definitionally equivalent to a natural number of the form \verb|suc n| but were nonetheless nonzero, it was difficult to write the fraction of \verb|+ 1 / k| instead of \verb|+ 1 / (suc k)|. This required a lot of additional work with a predecessor function, the result of which was not definitionally equivalent to \verb|+ 1 / k|. Thus, Lundfall's strategy was more difficult to use in practice.

Equality of real numbers is defined as a dependent data type, where the type constructor \verb|*≃*| can be used to construct a proof of \verb|x ≃ y|.

\begin{center}
\begin{BVerbatim}
data _≃_ : Rel ℝ Level.zero where
  *≃* : {x y : ℝ} → ((n : ℕ) {n≢0 : n ≢0} →
        ℚ.∣ seq x n ℚ.- seq y n ∣ ℚ.≤ (+ 2 / n) {n≢0}) →
        x ≃ y
\end{BVerbatim}
\end{center}

First, we note that the type \verb|Rel ℝ Level.zero| simply means that equality of real numbers is a relation on $\mathbb{R}$. Next, note that the real numbers \verb|x| and \verb|y| are implicit when calling the type constructor \verb|*≃*|. Given the proof that $|x_n - y_n| \le \frac{2}{n}$, which is also a parameter of \verb|*≃*|, along with the given goal (that we must prove \verb|x ≃ y|), Agda can deduce what it needs to write for the \verb|x| and \verb|y| parameters. Thus they are left implicit.

Lundfall's equality was not implemented as a data type. Instead, the type \verb|x ≃ y| was defined to be the proposition that $|x_n - y_n| \le \frac{2}{n}$ for all $n\in\mathbb{N}$, as follows \cite[p.~10]{lundfall}.

\begin{center}
\begin{BVerbatim}
_≃_ : Rel ℝ Level.zero
x ≃ y = (n : ℕ) →
        ℚ.∣ seq x n ℚ.- seq y n ∣ ℚ.≤ + 2 / (suc n)
\end{BVerbatim}
\end{center}  

There is the noted issue that the denominator of $\frac{2}{n}$ is written as \verb|suc n|. But more importantly, this definition is unfriendly towards implicit real number arguments. The function that proves reflexivity of equality is called \verb|≃-refl|. With our definition, given any real number \verb|x| and the goal \verb|x ≃ x|, we can write just \verb|≃-refl| to prove \verb|x ≃ x|, thus leaving the argument \verb|x| implicit. With Lundfall's definition, this is impossible. We must write the \verb|x| argument, thus writing \verb|≃-refl {x}|. This is a technical issue regarding dependent types defined as a data type (i.e., using the \verb|data| keyword) and defining types in the style of Lundfall. We will not go into the details, but being unable to leave \verb|x| implicit becomes extremely problematic when using the ring solver for the real numbers. Like our definition of the reals, our definition of equality is easier to use and is more maintainable in practice, even if the definition itself looks a little more complex.

When it comes to properties of the real numbers, such as the Archimedean property, my Agda formalization usually provides a normal and a ``fast'' version. For instance, the normal version of the Archimedean property is written \verb|archimedean-ℝ|, and the fast version is written \verb|fast-archimedean-ℝ|. The fast version is marked as \verb|abstract|, which, as mentioned in Section \ref{canonicity}, allows the user to manipulate the type of the marked function (corresponding to the Archimedean property, in this instance) without computing the function. This is a significant performance boost. On a Macbook Pro 2021 with an M1 Pro and 16GB of RAM, the library takes around a minute to typecheck when marking functions, like \verb|fast-archimedean-ℝ|, as abstract. Without marking them abstract, it takes several hours.

We have thus discussed some of the software engineering behind our library. There were many more decisions and problems similar to those we have discussed. But it would be far outside the scope of this thesis to discuss them all.

\section{Future Work}

A quick glance at \cite{bishopBridges} will show the reader that our library, at the time of writing, only covers somewhere between thirty and forty pages out of over four hundred. Thus, there is plenty of work to be done! Most importantly, the theories of differentiation and integration of functions of a real variable must be completed. 

There is also plenty of refactoring to do, as there is with any library. In particular, there is a lot of notation that was introduced later on in the project or modified that must be added to older work.

It may be interesting to modify the library to work in cubical Agda. However, there are definitions of the real numbers in homotopy type theory that are preferable to Bishop's, so such a project may ultimately be pointless.

The ring solver was immensely useful in this project. We wonder if a general inequality solver could be built. At the moment, every time the triangle inequality is used it must be directly cited, never left implicit or automatically used. Of course, analysts often implicitly use the triangle inequality in sequence with a number of other inequalities. Such an inequality solver would be immensely useful.

It is interesting to note that a general inequality solver exists in the Lean proof assistant \cite{lean}. Known as the \verb|linarith| function or ``tactic'', it assumes that the inequality to be proved is false and checks if it contradicts a true proposition in the context (e.g., a hypothesis). Obviously, the strategy of \verb|linarith| will not work constructively. We do not know if a constructive strategy exists. It is interesting that constructivism may, in some sense, be \textit{antithetical} to computation in this case.

Finally, now that a constructive theory of real numbers has been successfully implemented, it is high time for a classical theory of real numbers to be implemented. The extent to which problems like canonicity affect the classical reals should be studied, since it may make classical analysis in Agda untenable. Such a study may only be valuable to Agda's designers, since the Lean standard library contains virtually everything from undergraduate classical analysis along with a number of useful solvers (like \verb|linarith|) \cite{leanLibrary}. Once a classical analysis library is complete in Agda, it would be worth comparing it to the Lean implementation to see which is more maintainable and efficient, since proofs in each language tend to look extremely different, even if their content is the same. It may be deeply interesting to try and implement many of Lean's solvers in Agda, thus comparing their metaprogramming capabilities. It will be interesting to see how these two major proof assistants compete in the future.

\end{document}